\documentclass[preprint]{aastex}

\shorttitle{Planets around stars with a range of metallicities} 

\shortauthors{Ida and Lin}
\begin{document}

\title{The formation and retention of gas giant planets around
stars with a range of metallicities}

\author{Shigeru Ida}
\affil{Tokyo Institute of Technology,
Ookayama, Meguro-ku, Tokyo 152-8551, Japan}
\email{ida@geo.titech.ac.jp}

\and 

\author{D. N. C. Lin}
\affil{UCO/Lick Observatory, University of California, 
Santa Cruz, CA 95064}
\email{lin@ucolick.org}

\begin{abstract} 
The apparent dependence of detection frequency of extrasolar planets
on the metallicity of their host stars is investigated with Monte
Carlo simulations using a deterministic core-accretion planet
formation model.  According to this model, gas giants formed and
acquired their mass $M_{\rm p}$ through planetesimal coagulation
followed by the emergence of cores onto which gas is accreted.  These
protoplanets migrate and attain their asymptotic semi-major axis $a$
through their tidal interaction with their nascent disk.  Based on the
observed properties of protostellar disks, we generate $M_{\rm p}$-$a$
distribution. Our results reproduce the observed lack of planets with
intermediate mass $M_{\rm p} = 10$--100$M_{\oplus}$ and $a \la 3$AU
and with large mass $M_{\rm p} \ga 10^3 M_{\oplus}$ and $a \la 0.2$AU.
Based on the simulated $M_{\rm p}$-$a$ distributions, we also evaluate
the metallicity dependence of fraction of stars harboring planets that
are detectable with current radial velocity survey.  If protostellar
disks attain the same fraction of heavy elements which are contained
in their host stars, the detection probability around metal-rich stars
would be greatly enhanced because protoplanetary cores formed in them
can grow to several Earth masses prior to their depletion.  These
large masses are required for the cores to initiate rapid gas
accretion and to transform into giant planets.  The theoretically
extrapolated metallicity dependence is consistent with the
observation.  This correlation does not arise naturally in the
gravitational-instability scenario.  We also suggest other metallicity
dependence of the planet distributions that can be tested by on-going
observations.

\end{abstract}

\section{Introduction}

One of the most important observed characteristics of known extrasolar
planets is that their detection frequency $\eta_{\rm J}$ increases with the
metallicity $Z_\ast$ of their host stars \citep{Fischer,Santos04}.  Although
$Z_\ast$ could be significantly changed by the accretion of
planets/planetesimals onto the host stars after they have evolved onto
the main sequence, e.g., \citep{Sandquist98,Murray02}, the lack of any
$Z_\ast$ dispersion among members of stellar clusters 
\citep{Quillen02, Wilden02} indicate that the impact of this
process is very limited.  The $\eta_{\rm J}$-$Z_\ast$ correlation may also be
interpreted as evidence that the formation probability of gas giant
planets is greatly enhanced in metal-rich protostellar disks, in
accordance with the core-accretion scenario for formation of giant
planets, e.g., \citep{Mizuno80, BP86, P96, Ikoma00}.

In the conventional core-accretion scenario, heavy elements (metals)
in a protoplanetary disk condense into grains which form km-sized
planetesimals.  Through gravitational interaction and coagulation,
planetesimals evolve into cores through runaway \citep{Greenberg78,
WS89, ALP93, KI96} and oligarchic growth \citep{KI98,KI02}.  The
cores' eccentricity is effectively damped by their tidal interaction
with the ambient disk gas \citep{Artymowicz93, Ward93}.  After the
cores have swept up all residual planetesimals within a ``feeding-zone
width" $\sim 10$ Hill's radii \citep{KI98,KI02}, their growth in the
inner regions of the disks is stalled with an ``isolation'' mass
$M_{\rm c,iso}$.  If a core attains the critical mass, $M_{\rm c,acc}
\sim$ several $M_{\oplus}$, before the disk gas depletion, rapid gas
accretion onto it is initiated and it transforms into a gas giant
planet.  Since $M_{\rm c,iso}$ increases with the surface density
($\Sigma_{\rm d}$) of dusts (composed of metals) in protoplanetary
disks, formation of gas giants tends to be more prolific in a
metal-rich environment.

Here, using the deterministic planet formation model developed by
\citet{IL04} (hereafter referred to as Paper I), we calculate the
metallicity dependence of fraction ($\eta_{\rm J}$) of stars harboring
planets that are detectable with the current radial velocity surveys.
In \S2, we briefly describe our model.  In \S3, we simulate the
mass ($M_{\rm p}$) -- semimajor axis ($a$) distributions 
of extrasolar planets and their metallicity
dependence through a series of Monte Carlo simulations.  The
theoretically constructed distributions are consistent with observed
ones.  Using the simulated distributions, we evaluate $\eta_{\rm J}$.
Our model explains the observed metallicity dependence of $\eta_{\rm
J}$.  \S4 is summary and discussions.

\section{Model}

We carry out Monte Carlo simulations to reproduce the observed 
$M_{\rm p}$-$a$ distribution of extrasolar planets.  We basically follow the
methods of Paper I except slightly different choice of parameters.
Initial conditions of the Monte Carlo simulations are the $a$ of
planets and the dust and gas surface density 
($\Sigma_{\rm g}$ and $\Sigma_{\rm d} $) at $a$ in their nascent
disks.  For a given set of $\Sigma_{\rm g} (a)$ and $\Sigma_{\rm d}
(a)$, we numerically compute the entire formation and migration
sequence of individual cores and gas giant planets.  Here we briefly
summarize our prescription.  For details, see Paper I.

Since their observational data over the relevant length scales (a few
AU's) for planet-forming regions is not available, we scale their
quantitative values with global factors $f_{\rm d}$ and $f_{\rm g}$,
to those of the empirical minimum mass nebula (MMN) model for our
Solar system as
\begin{equation} 
\left\{ \begin{array}{ll} \Sigma_{\rm d} & = 10 \eta_{\rm ice} f_{\rm
d} (a/ {\rm 1 AU})^{-3/2} \;\; [{\rm g~cm}^{-2}], \\ \Sigma_{\rm g} &
= 2.4 \times 10^3 f_{\rm g} (a/ {\rm 1 AU})^{-3/2} \;\; [{\rm
g~cm}^{-2}],
\end{array} \right.  
\label{eq:sigma} 
\end{equation}
where the compositional scaling parameter $\eta_{\rm ice}=1$ inside
the ice condensation radius [$a_{\rm ice} = 2.7(M_\ast/M_\odot)^2$ AU
and 4.2 outside it \citep{Hayashi81}.
[Note that $\eta_{\rm ice}$ outside $a_{\rm ice}$ can be slightly 
smaller ($\sim 3.0$) \citep{Pollack94}.]
The mass of the host star
$M_\ast$ is scaled with that of the Sun $M_\odot$.  The dependence on
the power-law index for $\Sigma_{\rm d}$ and $\Sigma_{\rm g}$ are
examined in a separate paper.  To represent the decline in the
observational signatures of protostellar disks on timescales
$\tau_{\rm dep} \sim 1$--10 Myrs \citep{Haisch}, we assume exponential
decay of $\Sigma_{\rm g}$ as
\begin{equation} 
f_{\rm g} = f_{\rm g,0} \exp(- t/\tau_{\rm dep}),
\label{eq:f_g} 
\end{equation}
due to both viscous evolution and photoevaporation.  We neglect gas
replenishment onto the disk and any time variation in $f_{\rm d}$.  We
discuss the distributions of $f_{\rm d}$ and $f_{\rm g,0}$ below. 

With prescription (\ref{eq:sigma}), the cores' mass at time $t$ after
the formation of the disk is deduced to be (Paper I)
\begin{equation} 
M_{\rm c}(t) \simeq \left( \frac{\tau_{\rm dep}}
{4.8 \times 10^5{\rm years}} \right)^3 \eta_{\rm ice}^{3} 
f_{\rm d}^{3} f_{\rm g,0}^{6/5} \left(\frac{m}{10^{22}{\rm g}} \right)^{-2/5} 
\left( \frac{a}{1{\rm AU}} \right)^{-81/10}
\left(\frac{M_*}{M_{\odot}} \right)^{1/2} 
\left( 1 - \exp \left( {- 2t \over 5 \tau_{\rm dep}} \right)^3 \right) 
M_{\oplus}.
\label{eq:m_grow0} 
\end{equation} 
The dependence of $M_{\rm c}(t)$ on the mass $m$ of the field
planetesimals is due to the aerodynamical gas drag which regulates
their velocity dispersion.  Since e-folding growth time scale is
longer for larger $M_{\rm c}$, $M_{\rm c}(t)$ does not depend on
initial mass of the core.  In the nominal case, Eq.~(\ref{eq:m_grow0})
with $m=10^{22}$ g is used.  If $m$ is smaller or if radial migration
of cores due to their tidal interaction with gas \citep{GT80,Ward86}
and/or planetesimal disks \citep{Murray98,Ida00} occurs, core
accretion may be accelerated \citep{Inaba03,TI99,Rice03}.  We also carry out
the calculations with 3 times larger/smaller $dM_{\rm c}/dt$.
Note that we do not consider the change in $a$ of the cores
due to the radial migration, because the migration can be
a random walk \citep{Nelson04}.

Prior to severe depletion, the cores' growth is stalled when they
attain an isolation mass \citep{KI98, Kominami} which is given by
\begin{equation} 
M_{\rm c,iso}
\simeq 0.16 \eta_{\rm ice}^{3/2} f_{\rm d}^{3/2} \left( {a \over {\rm
1 AU}} \right)^{3/4} \left(\frac{M_*}{M_{\odot}} \right)^{-1/2}
M_\oplus.
\label{eq:isom} 
\end{equation} 
In our simulation, growth beyond the isolation after significant gas
depletion is taken into account (Paper I).  If radial migration of the
cores occurs, their growth barrier can be bypassed
\citep{TI99,Rice03} up to $M_{\rm c,noiso}$, which is given with
$M_{\rm c,no iso} \simeq \pi \Sigma_{\rm d} a^2$ (feeding zone width
$\Delta a \sim a$) by (Paper I)
\begin{equation} 
M_{\rm c,no iso} \simeq 1.2
\eta_{\rm ice}^{3/2} f_{\rm d} \left( {a \over {\rm 1 AU}}
\right)^{1/2}M_\oplus.  
\label{eq:noisom} 
\end{equation} 
In the nominal case,
the asymptotic masses of the cores are given by ${\rm min}(M_{\rm
c,iso},M_{\rm c,no iso})$.  We also examine tha case where $M_{\rm
c,no iso}$ is adopted in entire $f_{\rm d}$ and $a$.

For sufficiently rapid gas accretion to transform a core into a gas
giant planet, it must attain a critical core mass, $M_{\rm c,acc}
\sim$ several $M_{\oplus}$, before the disk gas depletion.  There are
at least two factors which limit the gas accretion rate.
Planetary atmosphere is no longer in hydrostatic equilibrium 
and gas accretion starts
when $M_{\rm c}$ exceeds $M_{\rm c,hydro}\simeq 
10 (\dot{M}_{\rm c}/10^{-6}M_{\oplus} {\rm yr}^{-1})^{0.25} M_{\oplus}$
\citep{Ikoma00}. 
\citet{P96} suggested that $M_{\rm c, hydro} > 10
M_\oplus$ because $\dot M_{\rm c}$ may be maintained at a modest
value, for a few Myr.  But, 
cores emerge with neighbors of comparable
masses.  Together they stir up the random motion of the residual
planetesimals, widen their feeding zone, and reduce their $\dot{M}_{\rm c}$.
In addition, the combined effects of gravitational scattering by
the cores and gas drag actually lead to a clearing of residual
planetesimals in the feeding zones of the cores \citep{TI97, Rafikov03} 
and a reduction in $M_{\rm c, hydro}$.
Thus, the onset of efficient gas accretion would occur
earlier than previously estimated.

With a negligible $\dot{M}_{\rm c}$, the gas accretion rate is
determined by the planet's total (core plus envelope) mass ($M_{\rm p}$) 
such that $dM_{\rm p}/dt = M_{\rm p}/\tau_{\rm KH}$ where
We adopt
\begin{equation} 
\tau_{\rm KH} = 10^{b} \left(\frac{M_{\rm p}}{M_\oplus} \right)^{-c} 
\; {\rm yrs},
\label{eq:t_KH}
\end{equation} 
with $b=10$ and $c=3$, which is a fitting formula for \citet{P96}'s
results with negligible planetesimal accretion.  
$\tau_{\rm KH}$ for $M_{\rm p} \sim 1$--$10 M_\oplus$ can be 
longer if modest planetesimal accretion induced by 
gas accretion is assumed [e.g., ``phase 2'' by \citet{P96}].  
But, the planetesimal accretion may be significantly reduced 
as discussed above.  In this case, phase 2 cannot be sustained.
On the other hand, if the core's migration is very fast,
the growth of a core does not decline until it acquires
mass $\sim M_{\rm c,noiso}$.  In this case, phase 2
does not exist either \citep{Alibert}. 
So, in the nominal case, we adopt $(b,c)=(10,3)$, 
but $(b,c)=(9,3)$ and $(11,3.5)$ are also examined
(the latter corresponds to the gas accretion including phase 2). 

The gas accretion rate increases rapidly with $M_{\rm p}$. Such
runaway gas accretion is terminated when the residual gas is depleted
globally on timescale $\tau_{\rm dep}$ or it
is severely depleted locally in the vicinity of their orbits 
due to gap opening, which is assumed to occur
if their Hill radius exceeds 1.5 times
the disk scale height (Paper I). 
When $M_{\rm p}$ becomes a significant fraction of 
$M_{\rm J}$ (Jupiter mass), 
planets' tidal torque on
the disk becomes larger than the viscous torque of the disk gas, 
resulting in a partial gap opening, e.g., \citep{Lubow}.
It reduces gas accretion rate.
Bondi accretion limit also becomes important in the reduction.
However, the elongated accretion timescale may be
still shorter than $\tau_{\rm KH}$ for  
$M_{\rm p} \sim$ several $M_{\oplus}$, 
so that the limited accretion at high $M_{\rm p}$ may not 
affect the mass distribution of gas giants significantly.
The critical mass $M_{\rm c,acc}$ for actual formation of gas giants
corresponds to the value of 
$M_{\rm p}$ for which $\tau_{\rm KH} \sim \tau_{\rm dep}$.  
Equation (\ref{eq:t_KH}) shows $M_{\rm c,acc} \sim$ several $M_{\oplus}$.
From eqs.~(\ref{eq:isom}) and (\ref{eq:m_grow0}), we find
that, for sufficiently large $f_{\rm d}$, it is possible for $M_{\rm
c}(\tau_{\rm dep}) > M_{\rm c,acc}$ and $M_{\rm c,iso} > M_{\rm
c,acc}$ so that gas giant can form readily.

After the gap opening, the planets undergo
``type-II" migration, in contrast to the cores' ``type-I'' migration
without a gap \citep{Ward97}.  In order to compute the protoplanets'
migration, we adopt the $\alpha$-prescription for effective viscosity
\citep{alpha} with a uniform $\alpha = 10^{-4}$ for all the disks such
that their viscous diffusion time scale ($\tau_{\nu}$) at $a \sim
10$AU is comparable to $\tau_{\rm dep} = 1$--$10$ Myrs.  The planets
migrate \citep{LP85,LP93} with a viscously evolving disk on time scales
(Paper I)
\begin{equation} 
\tau_{\rm mig} = \frac{a}{\dot{a}} = 10^6 f_{\rm g}^{-1} 
\left(\frac{M_{\rm p}}{M_{\rm J}} \right) 
\left(\frac{a}{1{\rm AU}}\right)^{1/2} \;{\rm yrs}.
\label{eq:tau_mig}
\end{equation} 
The planets' migration is terminated either when the disk is severely
depleted or when they reach the vicinity of their host stars.  We
artificially halt the planetary migration at 0.04AU.

Using these models, we carry out Monte Carlo simulations.
For simplicity, we generate a set of initial $a$'s of the
protoplanets, $\tau_{\rm dep}$ and $M_{\ast}$ with uniform
distributions in log scale in the ranges of 0.1--$100$AU,
$10^6$--$10^7$ yrs, and 0.7--$1.4M_\odot$, respectively.  
The assumed $a$ distribution corresponds to orbital
separations $\Delta a$ that is proportional to $a$, which may be
the simplest choice.
More refined choice of $\Delta a$ that takes into account
interactions between planets and detailed formation process
will be explored in a future paper.  
However, as long as $\Delta a$ increases with $a$ with
a moderate power index, calculated distributions of
gas giants may not change significantly.  

We discuss
the distribution of $f_{\rm d}$ and $f_{\rm g,0}$ in details, because
it is a key assumption of this paper.  A large range of dust mass in a
disk is inferred.  Observationally, the mass transfer rate from disks
onto classical T Tauri stars is $\sim 10^{-7}$--$10^{-9} M_\odot$
yr$^{-1}$ \citep{Calvet} with inferred disk masses equivalent to $0.1
\la f_{\rm g,0} \la 10$.  Independent mm observations \citep{BS96}
show a similar range of $f_{\rm d}$.  Here we assume that $\log_{10}
f_{\rm g,0}$ has a Gaussian distribution centered at 0 with the
dispersion $1.0$ and upper cut-off at $f_{\rm g,0} = 30$ (Paper I).

Since in the MMN, $f_{\rm d} = f_{\rm g,0} = 1$, the independently
determined mean values $\langle f_{\rm d} \rangle \sim \langle f_{\rm
g,0} \rangle \sim 1$ imply that the average value of metallicity in
the disk ($Z_{\rm disk}$) is $\sim Z_\odot$ (the Solar composition),
where [$Z_{\rm disk} \equiv {\rm log}_{10} (f_{\rm d} / f_{\rm
g,0}$)].  In Paper I, $f_{\rm d} = f_{\rm g,0}$ was assumed.  Here we
extend our calculation to a range of $Z_{\rm disk}$ in order to
account for the most pronounced correlation between the percentage of
stars with planets detectable with the current Doppler survey and
metallicity of their host stars ($Z_\ast$) \citep{Fischer,Santos04}, assuming
$Z_\ast \sim Z_{\rm disk}$.
From this inference, we explore the possibility that 
$\langle f_{\rm d} \rangle$ depends on $Z_\ast$ as 
$
\langle f_{\rm d}\rangle = 
10^{Z_{\rm disk} - Z_\odot} \sim 10^{Z_\ast-Z_\odot},
$
while $\langle f_{\rm g,0} \rangle$ is $\sim 1$, independent of
$Z_\ast$, since gas disk formation/evolution processes may not be
significantly affected by $Z_{\rm disk}$ $(\sim Z_\ast)$. We assume
$
f_{\rm d} = f_{\rm g,0} \times 10^{Z_\ast-Z_\odot}.
$
This estimate can be verified with additional direct measurement of
both gas and dust masses in protostellar disks in the near future.
All stellar
material is accreted through protostellar disks where differential
retention of gas and dust would affect the stellar metallicity.  These
key assumptions are supported by the remarkable
chemical homogeneity among young stars in the Pleiades cluster
\citep{Wilden02} and mature binary systems \citep{Desidera04}.

\section{Numerical Results}

The simulated $M_{\rm p}$--$a$ distributions with $\Delta Z_\ast
\equiv Z_\ast-Z_\odot = 0$ are shown in Figure 1a.
This is the nominal case where $m=10^{22}$ g, $M_{\rm c,iso}$ and
$\tau_{\rm KH}=10^{10} (M_{\rm p}/M_\oplus)^{-3}$ yrs are adopted.
For comparison, the observed distributions are also plotted in Figure 1b.
The observational limits in the current Doppler survey, $M_{\rm p} \ga
100 (a/1{\rm AU})^{1/2}M_{\oplus}$ and $a \la 3$AU (around F, G, K
dwarfs within $\sim 50$ pc) are marked by dotted lines.
A deficit of planets with an asymptotic mass
$M_{\rm p} = 10$--100$M_{\oplus}$ and $a \la 3$AU (within a ``planet
desert") is apparent in Fig.~1a.  This sparsely populated region 
is primarily
caused by the runaway gas accretion onto cores with $M_{\rm p} \ga$
several $M_\oplus$ (Paper I).  This region divides the terrestrial
(rocky) planet, gas and ice giant domains in the planets' mass-period
distribution.  Planetary migration generally sharpens the boundaries
of the domains (Paper I; \citet{Udry}).
Since migration is slower for larger $M_{\rm p}$
(Eq.~[\ref{eq:tau_mig}]), relatively massive planets are less likely
to migrate to inner regions, resulting in a deficit of planets with
$M_{\rm p} \ga 10^3 M_{\oplus}$ at $a \la 0.2$AU (Fig.~1a;
also see discussion by \citet{Udry}).  If larger $\alpha$ is adopted,
$\tau_{\nu}$ becomes short compared with $\tau_{\rm dep}$, so that
even relatively massive planets can migrate significantly, resulting
in less clear deficit.  
We found that the two planet-depleted regions exist
for other $\Delta Z_*$ as well, because the ``planet desert'' and 
the other are determined by the core accretion model
for gas giants and $\tau_{\rm dep}/\tau_\nu$, respectively.
The observed distributions may exhibit the above two deficits and are
consistent with our theoretical models with $\tau_{\rm dep} \sim
\tau_\nu$ at $\sim 10$ AU.  
Adopting the same condition, the simulated $a$-distribution 
of the gas giant planets can match 
with the observation [Figure 1c; also see
\citet{Armitage02} and \citet{Trilling02}].  

The 0.04 AU cut-off in the planets' $a$ distribution in our
results may also be less abrupt and more consistent with the observed
distributions if we adopt more relaxed inner boundary conditions.
Many of giant planets may have migrated toward the vicinity of their
host stars \citep{Lin96}.  The theoretically determined ratio of the
giant planets with $a < 0.06$AU (hot Jupiters) to those between
0.2--2AU is $\sim 1$--10 for stars with $Z_\ast \sim Z_\odot$, which
is order of magnitude larger than its observed value ($\sim 0.2$).
This discrepancy suggests that more than $90\%$ of the hot Jupiters
which migrated to the stellar vicinity are either consumed
\citep{Sandquist98} or tidally disrupted \citep{Gu} by their host
stars.  Provided these disruptive events occur before the host stars
become main sequence stars with relatively shallow convection zones,
their apparent chemical homogeneity may be preserved to match the
observation of the Pleiades cluster \citep{Wilden02}.  As the
migrating protoplanets sweep planetesimals along their paths, this
process may self-regulate the amount of residual heavy elements in the
disk.

Figures 1c and d show that the normalized slopes of the $a$ and 
$M_{\rm p}$ distributions are independent of $\Delta Z_\ast$, 
because they are
determined by $\tau_{\rm dep}/\tau_\nu$ and the core accretion model
for gas giants, respectively.
In these figures, we do not include planets stalled 
at 0.04AU, since a large fraction of them
may plunge into their host stars.

Although the normalized distributions are independent of
$\Delta Z_\ast$,  we found that 
formation and retention rates of gas giant planets 
rapidly increase with $\Delta Z_\ast$.
In order to account for the $\Delta Z_\ast$-dependence in 
the $M_{\rm p}$ and $a$ distribution, we first describe how planet formation
depends on $f_{\rm d}$ [for details, see Paper I and \citet{KI02}].
In a disk similar to the MMN ($f_{\rm d} = f_{\rm g,0} = 1$), for $a <
a_{\rm ice}$, the core growth is limited to $M_{\rm c,iso} < M_\oplus$
where $M_\oplus$ is the Earth mass.  Although growth beyond isolation
(up to $\sim M_\oplus$) is possible for terrestrial planets, it is
attainable only after severe gas depletion \citep{Kominami}.  At these
regions, both $M_{\rm c,iso}$ and $M_{\rm c,noiso}$ are $<M_{\rm c,
acc}$ such that the cores cannot acquire a massive gaseous envelope.
At the outer regions where $a \gg a_{\rm ice}$, although $M_{\rm
c,iso} > M_{\rm c,acc}$, the timescale required for $M_{\rm c}$ to
reach $M_{\rm c,acc}$ is $\gg \tau_{\rm dep}$.  The cores evolve
into ice giants, similar to Uranus and Neptune.  In the disk similar
to th MMN, gas giants can form only slightly outside the ice boundary,
because the cores may acquire $M_{\rm c,acc}$ there within a few Myr
and enter a rapid gas accretion phase prior to the global disk
depletion.  Gas giants can form inside $a_{\rm ice}$ in some massive
disks (with $f_{\rm d} \ga 5$).  Rocky cores are formed with $M_{\rm
c,iso} \la M_\oplus$ in the inner regions and ice giants are formed
with $M_{\rm c} (\tau_{\rm dep}) \la M_\oplus$ in the outer regions of
low-mass disks with $f_{\rm d} \la 0.5$, without formation of gas
giants.

Around metal-deficient stars, formation efficiency of gas giants is
low because disks with sufficiently large $f_{\rm d}$ are only at the
tail end of the distribution.  Around the metal-rich stars, the
efficiency is high because both $dM_{\rm c}/dt$ and $M_{\rm c,iso}$
increase rapidly with $f_{\rm d}$ (eqs.~\ref{eq:m_grow0} and
\ref{eq:isom}) and $M_{\rm c,iso} \ga M_{\rm c,acc}$ even interior to
$a_{\rm ice}$.  These account for the $\Delta Z_\ast$-dependence.
In order to directly compare with the observed frequency of extrasolar
planets, we plot in Figures 2 the theoretical determination for fraction
($\eta_{\rm J}$) of stars which bear giant planets detectable with the
current Doppler survey, as a function of their metallicity.  Here, the
distribution of $f_{\rm d}$ is the same as that in Figures 1.  For
each $f_{\rm d}$, $a$ is selected as $\log (a_{j+1}/a_j) = 0.2$
$(j=1,2,...)$.  
We do not include planets stalled at 0.04AU in the evaluation of
$\eta_{\rm J}$.

These results indicate $\eta_{\rm J}$ increases linearly with $\Delta
Z_\ast$, which is consistent with observational results of
\citet{Fischer} (open circles in Figures 2).  As shown in Figure 2a,
$\eta_{\rm J}$ does not depend on the choice of $\tau_{\rm KH}$
[$(b,c)=(9,3), (10,3)$ or $(11,3.5)$; $\tau_{\rm KH} = 10^{b} (M_{\rm
p}/M_\oplus)^{-c}$] nor that of asymptotic core mass [min$(M_{\rm
c,iso},M_{\rm c,no iso})$ or $M_{\rm c,no iso}$].  The reasons are as
follows.  As long as $\tau_{\rm KH}$ has strong dependence on $M_{\rm
p}$, even significant change in $b$ results in relatively small change
of $M_{\rm c,acc}$, 
the core mass that enables rapid gas accretion prior to disk gas
depletion.  Since in our simulations, most gas giant planets are
formed with such parameter ranges where $f_{\rm d}$ and $a$ are 
relatively large and $M_{\rm c, iso} \ga M_{\rm c,no iso}$, 
the distributions hardly change even if $M_{\rm c,no iso}$
is adopted in entire $f_{\rm d}$ and $a$ (although it affects the
distributions of terrestrial planets [Paper I]).

Although the absolute values of $\eta_{\rm J}$ are higher/lower for
faster/slower core accretion, the dependence on $\Delta Z_\ast$ is
similar (Figure 2b).  If disk depletion is caused by viscous
diffusion, the assumption that $\tau_{\rm dep} \sim \tau_\nu$ at $\sim
10$ AU is reasonable.  However, the observation of dense clusters
suggest that gas depletion in protostellar disks is primarily driven
by photo-evaporation \citep{Johnstone} in these clusters, where a few
massive stars contribute to nearly the entire UV ionizing flux.  In
order to consider the possibility of rapid gas depletion in a dense
cluster, we carried out the additional simulations with $\tau_{\rm
dep} = 10^5$--$10^6$ yrs. In this case, formation rates of giant
planets are lower, resulting in smaller $\eta_{\rm J}$ (Figure 2b).
But, the dependence on $\Delta Z_\ast$ does not change also in this
case.  This general agreement is a generic feature of the core
accretion scenario and the assumption for $f_{\rm d}$ and
$f_{\rm g,0}$. It may
not arise naturally from the gravitational instability scenario
\citep{Boss97}.

\section{Discussion}

Through Monte Carlo simulation using the deterministic planet
formation model developed in Paper I, we find that the formation
probability of gas giant planets increases rapidly with the
metallicity of their host stars.  Our model quantitatively accounts
for observational correlation between stellar metallicity and the
fraction of stars harboring giant planets that are detectable with the
current Doppler survey.  Our model also explains the two
observationally suggested planet-depleted regions in $M_{\rm p}$-$a$
distributions of extrasolar planets: the deficit in intermediate mass
$M_{\rm p} = 10$--100$M_{\oplus}$ and $a \la 3$AU, and that in large
mass $M_{\rm p} \ga 10^3 M_{\oplus}$ and $a \la 0.2$AU.

Our model can be tested by more detailed observation.  We 
show that the normalized slope of the $a$ and $M_{\rm p}$ 
distributions of gas giants and
the planet-depleted regions are independent of $\Delta Z_\ast$,
because these are regulated by $\tau_{\rm dep}/\tau_\nu$ and the core
accretion model.  The lower limit mass $M_{\rm p,lim}$ of hot Jupiters
depend on $\Delta Z_\ast$ as $M_{\rm p,lim} \propto 10^{-(3/2)\Delta
Z_\ast}$ (Paper I).  The outer boundary radius of the distribution of
gas giants $a_{\rm out} \propto 10^{(10/27)\Delta Z_\ast}$ (Paper I),
although interactions among gas giants, which are not taken into
account in the present paper, may make the boundary less clear
\citep{MW02}. These predictions of $\Delta Z_\ast$-dependence can be
tested by statistics of extrasolar planets with increased detection
number in a few years.

Our results provide supporting evidences for the core-accretion
scenario and they suggest that the detection of gas giant planets may
be used to infer the presence of rocky planets around the same host
stars.  "Habitable planets" which can retain liquid water on their
surface \citep{Kasting} may have $0.1M_\oplus \la M_{\rm p} \la
10M_\oplus$ and $a\sim$ 0.8--1.5$\times (M_\ast/M_\odot)^2$ AU.  The
results in Figure 1 indicate that the frequency of "habitable planets"
is comparable to or more than that of detectable giant planets (at
present, more than 100 planets have been detected around F, G, K dwarfs
within 50 pc), although some fraction of the habitable planets may be
destabilized by migrating giant planets.

{}

\vspace{1em} 
\noindent ACKNOWLEDGMENTS.  
We thank Debra Fischer, Greg Laughlin and Masahiro Ikoma for useful
discussions and the anonymous referee for helpful comments.  
This work is supported by the NASA Origins program, JPL
SIM program, NSF and JSPS.

\vspace{1em}
\noindent CORRESPONDENCE should be addressed to S. I. (ida@geo.titech.ac.jp).

\clearpage
\vspace{1em}
\noindent FIGURE LEGENDS

\noindent {\bf Fig.1}

Theoretically predicted mass ($M_{\rm p}$) and semi-major axis ($a$)
distributions of extrasolar planets.
For parameters, see text. 
In panel (a), the computed result with $\Delta Z_* =0$ is shown.
Totally 20000 planets are calculated.
For comparison, observational data are plotted in panel (b).
(Since orbital inclinations are unknown, minimum masses are plotted).  
The observational limits in the current Doppler survey
are marked by dotted lines.
One dimensional normalized distributions of observable planets is plotted
as a function of $a$ (panel c) and $M_{\rm p}$ (panel d).
Large open circles are observed results.  Small filled circles, pentagons, 
squares, triangles, and crosses express
the theoretical results with $\Delta Z_* = 0.5, 0.25, 0, -0.25$ and $-0.5$,
respectively.
Here 1000 observable planets are used in each $\Delta Z_*$ result.

\vspace{1em}
\noindent {\bf Fig.2}

The fraction ($\eta_{\rm J}$) of stars which bears giant planets 
currently detectable with Doppler survey, as a function of their
metallicity ($\Delta Z_*$). 
Large open circles are observational results \citep{Fischer}. 
Other small symbols with lines represent
theoretical predictions.  Filled circles express the nominal case
where core accretion is given by Eq.~(\ref{eq:m_grow0}) with $m=10^{22}$ g,
the asymptotic masses are ${\rm min}(M_{\rm c,iso},M_{\rm c,no iso})$,
and $\tau_{\rm KH}=10^{b} (M_{\rm p}/M_\oplus)^{-c}$ yrs with $(b,c)=(10,3)$.  
In panel (a), crosses, filled triangles and squares express 
the cases with $M_{\rm c,no iso}$, $(b,c)=(9,3)$, and $(b,c)=(11,3.5)$,
respectively.
In panel (b), crosses, filled triangles and squares express 
the cases with 3 times faster core accretion rate, 
3 times slower one, and $\tau_{\rm dep} = 10^5$--$10^6$ yrs, respectively.

\clearpage
\begin{figure}
\plotone{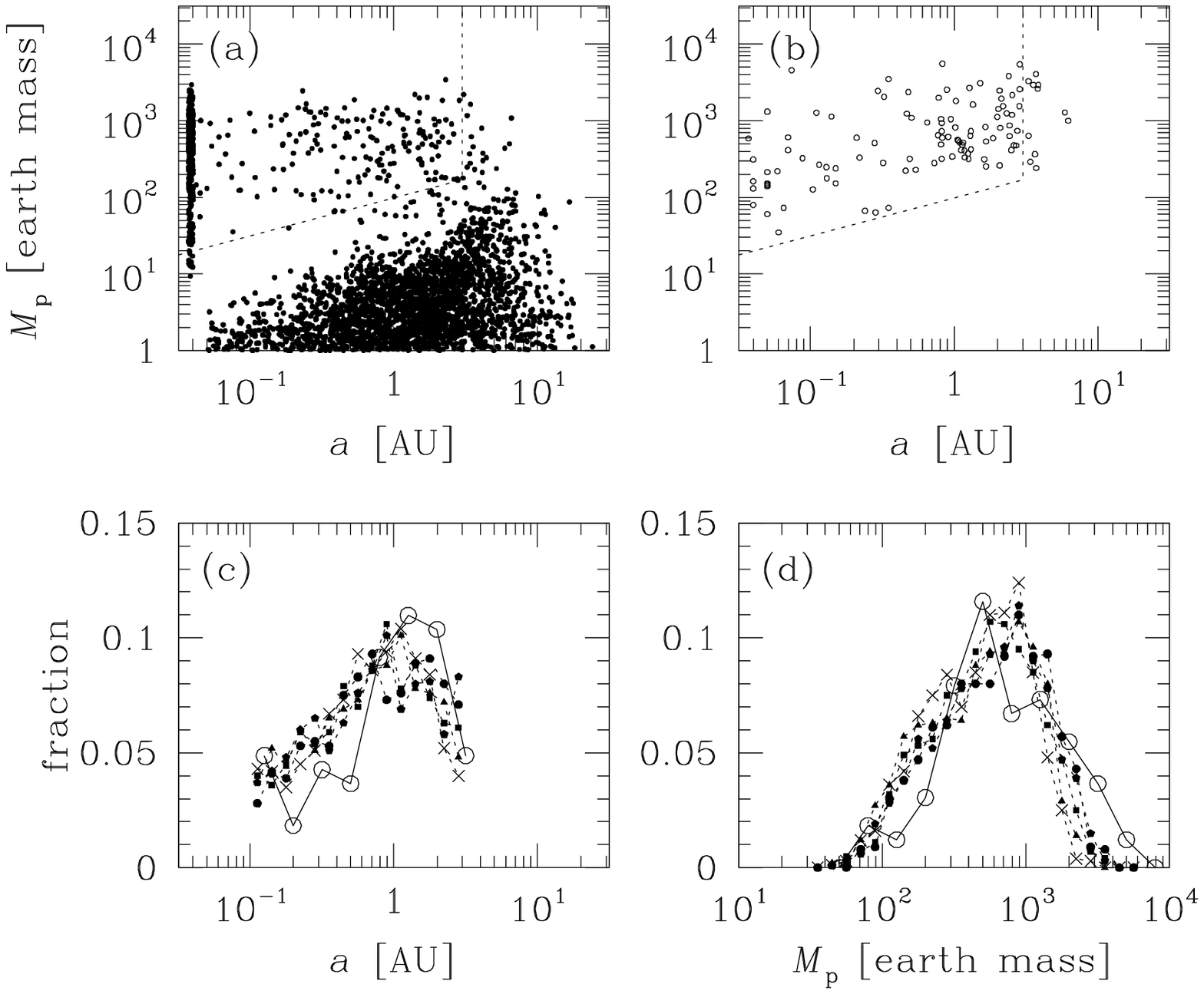}
\caption{   }
\label{fig:1}
\end{figure}

\clearpage
\begin{figure}
\plotone{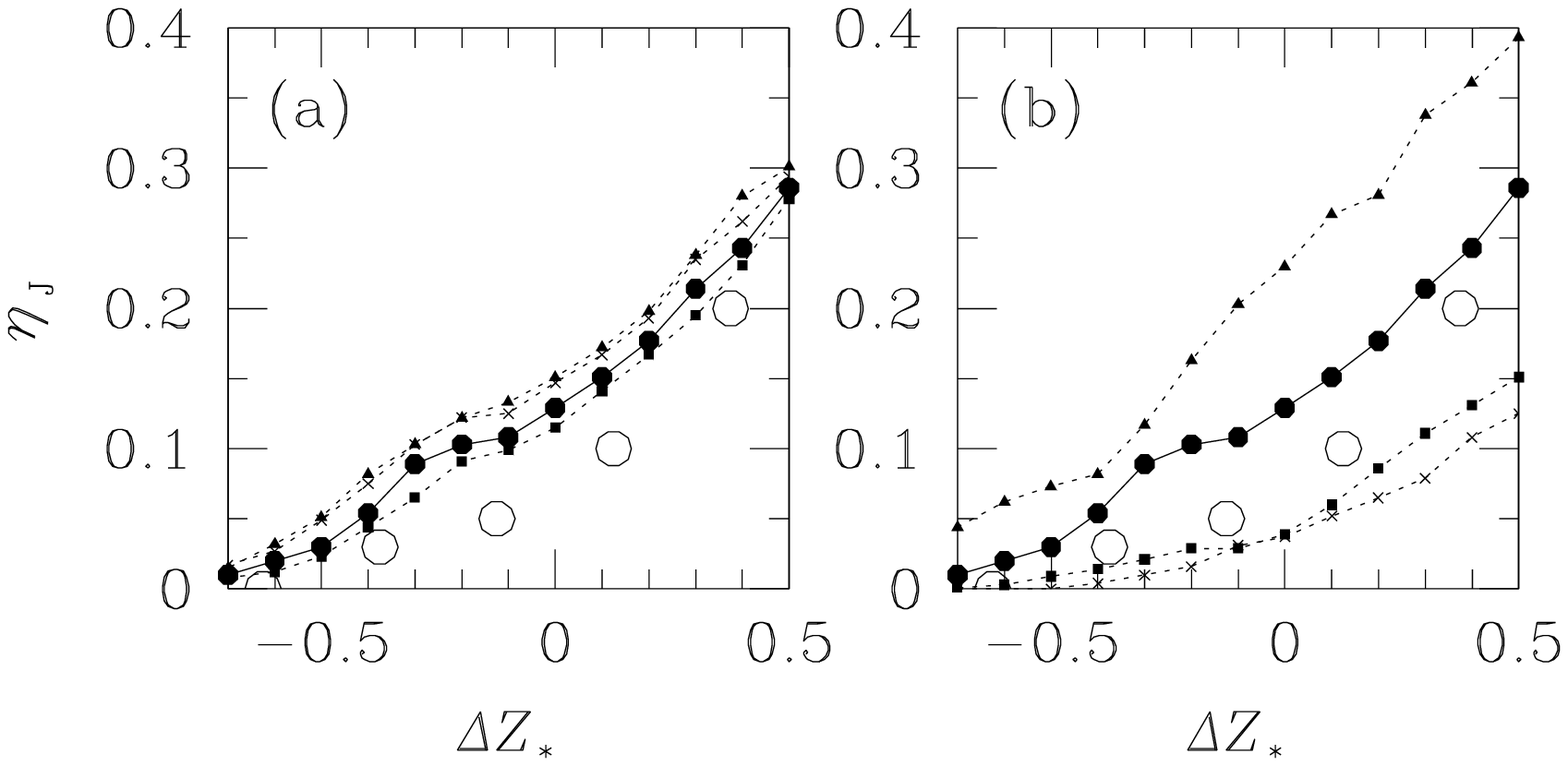}
\caption{   }
\label{fig:2}
\end{figure}

\end{document}